\documentclass{llncs}
\usepackage{amssymb}
\usepackage{amsfonts}
\usepackage{subcaption}
\usepackage[cmex10]{amsmath}
\usepackage[T1]{fontenc}
\usepackage{mathtools} 
\usepackage{siunitx}
\usepackage{hyperref}
\usepackage{graphicx}
\usepackage{float}
\usepackage[utf8]{inputenc}
\usepackage[english]{babel}
\usepackage{authblk}
\usepackage{colortbl}
\usepackage{ctable}
\usepackage{url}
\usepackage{multirow}

\begin{document}

\title{Transfer Learning from Partial Annotations for Whole Brain Segmentation}
 \author{Chengliang Dai\inst{1} \and
 Yuanhan Mo\inst{1} \and 
 Elsa Angelini\inst{2} \and
 Yike Guo\inst{1} \and
 Wenjia Bai\inst{1,3}}

 \institute{Data Science Institute, Imperial College London, London, UK
 \and
 ITMAT Data Science Group, Imperial College London, London, UK
 \and
 Department of Brain Sciences, Imperial College London, London, UK
 }

\maketitle
\begin{abstract}
Brain MR image segmentation is a key task in neuroimaging studies. It is commonly conducted using standard computational tools, such as FSL, SPM, multi-atlas segmentation etc, which are often registration-based and suffer from expensive computation cost. Recently, there is an increased interest using deep neural networks for brain image segmentation, which have demonstrated advantages in both speed and performance. However, neural networks-based approaches normally require a large amount of manual annotations for optimising the massive amount of network parameters. For 3D networks used in volumetric image segmentation, this has become a particular challenge, as a 3D network consists of many more parameters compared to its 2D counterpart. Manual annotation of 3D brain images is extremely time-consuming and requires extensive involvement of trained experts. To address the challenge with limited manual annotations, here we propose a novel multi-task learning framework for brain image segmentation, which utilises a large amount of automatically generated partial annotations together with a small set of manually created full annotations for network training. Our method yields a high performance comparable to state-of-the-art methods for whole brain segmentation. 
\noindent 
\end{abstract}

\section{Introduction}
Magnetic resonance imaging (MRI) plays an important role in human brain studies due to its good performance on presenting anatomy, pathology and function of the brain. Accurate segmentation of brain MRI scans is a prerequisite for measuring volume, thickness and shape of brain structure, which allows researchers to track and study the development, ageing and diseases of the brain \cite{douaud2014common}. Brain image segmentation is a time-consuming process when conducted manually, which typically takes several hours for a single subject. Therefore computational tools including FSL \cite{jenkinson2012fsl}, SPM \cite{ashburner2000voxel}, MALP-EM \cite{ledig2015robust} etc have been developed to automatically segment brain MRI scans and to enable large-scale population-based imaging studies. Most of these computational tools segment the scans by performing linear and nonlinear registration between a manually annotated brain atlas and a target scan and then propagating the atlas. Despite the efficiency they bring, these tools still suffer problems such as expensive computational cost and potential failures in image registration. Furthermore, strict pre-processing steps including brain stripping and bias correction are required to improve the reliability of these computational tools. 

Neural networks have been explored and widely used for brain segmentation in recent years. Comparing to conventional brain image segmentation pipelines that are registration-based, network-based methods use pairs of images and manual annotations to train a discriminative model for inferring the segmentation of a new scan. Such difference brings a few advantages: (i) pre-processing can be potentially simplified \cite{rajchl2018neuronet}; (ii) processing time is significantly reduced without sacrificing the segmentation accuracy. Segmenting brain with network-based models also has drawbacks as these models require massive amount of annotated data for model training. The limited amount of annotations for brain images has become one of the biggest challenges for applying neural networks to brain image segmentation. 

Previous works have been exploring ways for training image segmentation networks with limited annotations. A common approach is to fine-tune a pre-trained network from large image datasets like ImageNet \cite{tajbakhsh2016convolutional}. In \cite{roy2019quicknat}, an encoder-decoder model is pre-trained with auxiliary labels generated by FreeSurfer and then fine-tuned with an error corrective boosting loss. In \cite{moeskops2016deep}, a multi-task image segmentation model is investigated to learn features that can be shared between MRI scans of different parts of human body. Generative adversarial networks (GANs) are adopted in \cite{shin2018medical} for data augmentation, which indicates a better performance than conventional augmentation methods. 

Here we propose a novel brain image segmentation network, which leverages a massive set of automatically generated partial annotations (sub-cortical segmentations from FSL) for network pre-training and then perform transfer learning onto a small set of full annotations (manual whole brain segmentations). Compared to \cite{roy2019quicknat}, our method is conducted in 3D but with less convolutional layers. We demonstrate how features learnt from partial annotations in the source domain can be adapted to the target domain. With very limited annotations, our method achieves a performance comparable to state-of-the-art methods for brain image segmentation.

\begin{figure}[h]
\begin{center}
  \includegraphics[scale=0.35]{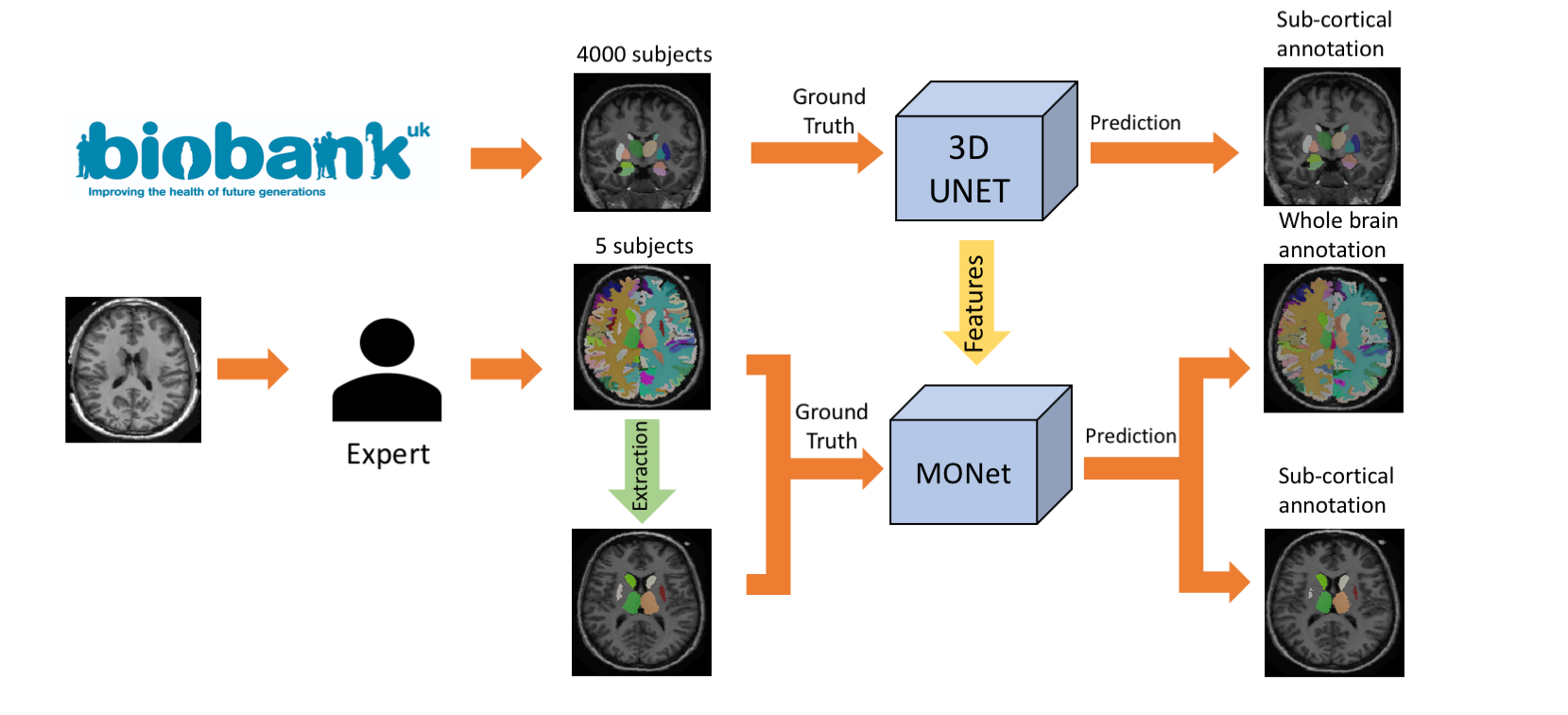}
  \caption{Two-stage training scheme: Stage 1: pre-training; Stage 2: joint training.}\label{2stage}
  \end{center}
\end{figure}

\section{Method}
\label{sec:method}
Our work adopts a two-stage training scheme as illustrated in Fig.~\ref{2stage}. Stage 1 pre-trains the segmentation network using a large set of automatically generated partial annotations. Stage 2 fine-tunes the network by jointly training on partial annotations and a small set of full annotations.

\subsection{Pre-training with partial annotations}
In this work, partial annotation refers to segmentation that only covers part of the brain structures. In our case, it refers to segmentation of 15 sub-cortical structures automatically generated by FSL. Full annotation refers to segmentation of whole brain structures manually annotated by human experts, which is a superset of partial annotation and consists of 138 structures. Since partial annotations are automatically generated, it is easy to acquire many of them. On the other hand, acquiring full annotations is more difficult as it requires extensive manual labour. 

A 3D U-Net is employed for pre-training on partial annotations, using categorical cross-entropy as the loss function,
\begin{equation} 
\mathcal{L} = -\sum_{v}g_{l}^{w}(v)\log p_{l}^{w}(v)
\end{equation}
where \(p_{l}^{w}(v)\) is the the predictive probability of partial segmentation belonging to class \(l\) at voxel \(v\) and \(g^{w}(v)\) is the probability of it belonging to its actual class.

\begin{figure}[h]
\begin{center}
  \includegraphics[scale=0.33]{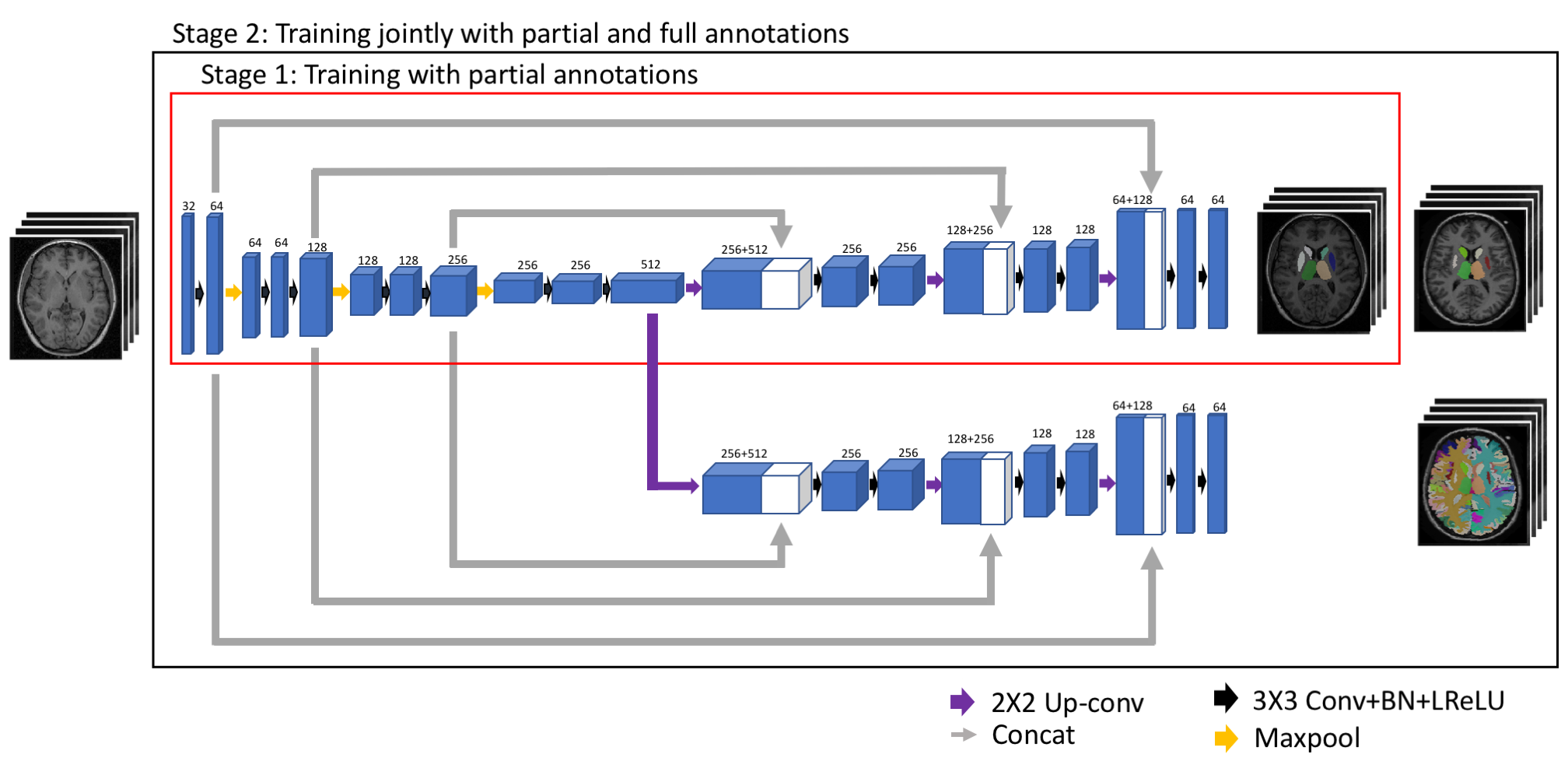}
  \caption{Network architectures used in stage one and two training.}\label{arc}
  \end{center}
\end{figure}

\subsection{Joint training with full annotations}
We employ a multi-task learning framework for the second stage. The encoder is consistent with the architecture used in the first stage. Two decoders are used, so that the two tasks (partial segmentation and full segmentation) can be jointly trained. We refer our method as the multi-output network (MO-Net). The encoder and both decoders are loaded with the pre-trained parameters. Multi-output design encourages the encoder to learn shared features for partial segmentation and full segmentation. The partial segmentation used for joint training is extracted from the full segmentation, which are manual segmentations of the whole brain. Since manual segmentations have always been considered as 'gold standard' and should be more reliable than segmentations from automatic tools, the trained MO-Net should also be able to provide more accurate partial segmentation than the one trained in the first stage. The multi-output design given in Fig.~\ref{arc} is similar to the one described in \cite{rajchl2018neuronet}, which allows the network to learn jointly from two segmentation maps in order to achieve more accurate prediction and to have the potential to provide segmentation output for various annotation protocols. However, the difference is that we use a modified U-Net instead of ResNet and FCN adopted in \cite{rajchl2018neuronet}, and our network is loaded with the parameters learnt from the pre-training stage.

A weighted loss that combines the overall loss of two decoders of MO-Net for joint training is formulated as,
\begin{equation}
    \mathcal{L}_{MO-Net} = -\sum_{v}\lambda_{s}g^{s}(v)\log p_{m}^{s}(v) - \sum_{v}\lambda_{w}g^{w}(v)\log p_{l}^{w}(v)
\end{equation}
where \(p_{m}^{s}(v)\) is the the predictive probability of full segmentation belonging to class \(m\) at voxel \(v\) and \(g^{s}(v)\) is the probability of it belonging to its actual class. \(\lambda_{s}\) and \(\lambda_{w}\) are the weights for overall loss function. To balance between the learning tasks for partial segmentation and full segmentation, we assign 0.5 to both losses in the overall loss function.

\section{Experiments and Results}
\label{sec:EAR}
\subsection{Datasets}
\textbf{UK Biobank Dataset (UKBB)} 4,000 MRI brain scans from the UK Biobank are used. Automatic sub-cortical segmentations of 15 regions by FSL are used as partial annotations for pre-training.

\noindent\textbf{Hammers Adult Atlases (HAA)} The HAA dataset \cite{hammers2003three}\cite{gousias2008automatic} contains brain atlases for 20 subjects with manual annotations for 67 regions. The dataset is split into 5/2/13 for training, validation and test.

\noindent\textbf{MICCAI 2012 Multi-atlas Labelling Challenge (MALC)} The MALC dataset \cite{landman2012miccai} contains MRI scans from 30 subjects (15 subjects for training) with manual annotations for the whole brain for 138 regions and 132 regions are used for performance evaluation. The dataset also includes 5 follow-up scans, but they are excluded in our work. The dataset is split into 15/2/13 for training, validation and testing.

The manual annotations from the HAA and MALC datasets are regarded as ground truth in evaluation.

\subsection{Preprocessing and Training}
The typical brain image resolution is \(256^{3}\), with isotropic spatial resolution of \(1mm^{3}\). All images were rigidly registered to MNI space and normalized to zero mean and unit standard deviation. For training the network, 3D patches of size \(128^{3}\) were randomly drawn from the brain images. Batch size was set to 1 due to the limitation of GPU memory. Random elastic deformation was applied to the 3D patches for data augmentation. Cropping and augmentation were performed on-the-fly. Adam optimiser with a starting learning rate of 0.001 was used for both stages of network training. Leaky rectified linear unit (LeakyReLU) with a negative slope of 0.01 is applied as the activation function. For the proposed method, pre-training was ran for 3 epochs and joint training was ran for 200 epochs. We also trained a standard U-Net as a baseline method for comparison.

\subsection{Results}
We evaluated the performance of MO-Net in terms of Dice score. For comparison, two versions of U-Nets were trained, one trained from scratch (U-Net (FS)) and the other fine-tuned (U-Net (FT)) on MALC and HAA respectively. For evaluating whole brain segmentation performance on MALC, we also compared our result to SLANT8 and SLANT27 \cite{huo20193d}, which is based on fine-tuning 8 and 27 3D U-Nets pre-trained with 5111 subjects for different locations of brain.

As shown in Tab.~\ref{tab:malc} and Tab.~\ref{tab:haa}, our method outperformed the U-Net trained from scratch by 26\% on MALC dataset and 19\% on HAA dataset. MO-Net also shows slight improvements over fine-tuned U-Net, SLANT8 and SLANT27 on both MALC and HAA datasets. We further compared to QuickNAT \cite{roy2019quicknat} on the same 25 brain structures as in their paper on the MALC dataset. The result is given in Tab.~\ref{tab:malc25}. MO-Net outperformed the fine-tuned U-Net, SLANT8 and SLANT27 by a small margin, although the performance is inferior to QuickNAT.

\begin{table}[h]
\parbox{.45\linewidth}{
\centering
\begin{tabular}{cc}
\hline
Method & Dice (mean$\pm$std) \\ \hline
U-Net (FS) & 0.623$\pm$0.095 \\ 
U-Net (FT) & 0.782$\pm$0.043 \\
SLANT8 \cite{huo20193d} & 0.768$\pm$0.011 \\ 
SLANT27 \cite{huo20193d} & 0.776$\pm$0.012 \\ 
MO-Net     & \textbf{0.785}$\pm$\textbf{0.070} \\ \hline
\end{tabular}
\caption{Whole brain segmentation accuracy on MALC. \label{tab:malc}}
}
\hfill
\parbox{.45\linewidth}{
\centering
\begin{tabular}{cc}
\hline
Method & Dice (mean$\pm$std)                          \\ \hline
U-Net (FS) & 0.706$\pm$0.032 \\
U-Net (FT) & 0.821$\pm$0.019 \\
MO-Net     & \textbf{0.843}$\pm$\textbf{0.037} \\ \hline
\end{tabular}
\caption{Whole brain segmentation accuracy on HAA. \label{tab:haa}}
}
\end{table}

\begin{table}[h]
\centering
\begin{tabular}{cc}
\hline
Method & Dice (mean$\pm$std) \\ \hline
U-Net (FS) & 0.775$\pm$0.035 \\
U-Net (FT) & 0.809$\pm$0.021\\
SLANT9 \cite{huo20193d}  & 0.817$\pm$0.036 \\
SLANT27 \cite{huo20193d}  & 0.823$\pm$0.037 \\
QuickNAT \cite{roy2019quicknat} & 0.901$\pm$0.045 \\
MO-Net     & 0.838$\pm$0.049 \\ \hline
\end{tabular}
\caption{Segmentation accuracy for 25 structures on MALC. \label{tab:malc25}}
\end{table}

\begin{table}
\parbox{.45\linewidth}{
\centering
\begin{tabular}{cc}
\hline
Method     & Dice (mean$\pm$std) \\ \hline
U-Net (FS) & 0.649$\pm$0.145           \\
U-Net (FT) & 0.835$\pm$0.062           \\
FSL        & 0.637(9 failed)$\pm$0.216 \\
MO-Net     & 0.826$\pm$0.029           \\ \hline
\end{tabular}
\caption{Segmentation accuracy for 15 sub-cortical structures on MALC. \label{tab:malc_sub}}
}
\hfill
\parbox{.45\linewidth}{
\centering
\begin{tabular}{cc}
\hline
Method     & Dice (mean$\pm$std)  \\ \hline
U-Net (FS) & 0.612$\pm$0.103 \\
U-Net (FT) & 0.874$\pm$0.053 \\
FSL        & 0.763$\pm$0.043 \\
MO-Net     & \textbf{0.879}$\pm$\textbf{0.091} \\ \hline
\end{tabular}
\caption{Segmentation accuracy for 15 sub-cortical structures on HAA. \label{tab:haa_sub}}
}
\end{table}

For sub-cortical segmentation, we compared our result to FSL FIRST and U-Net. The proposed method MO-Net shows similar Dice score performance to fine-tuned U-Net and it is better than FSL and U-Net trained from scratch. The result is shown in Tab.~\ref{tab:malc_sub} and Tab.~\ref{tab:malc}.

\begin{figure}
\begin{center}
  \includegraphics[width=10cm]{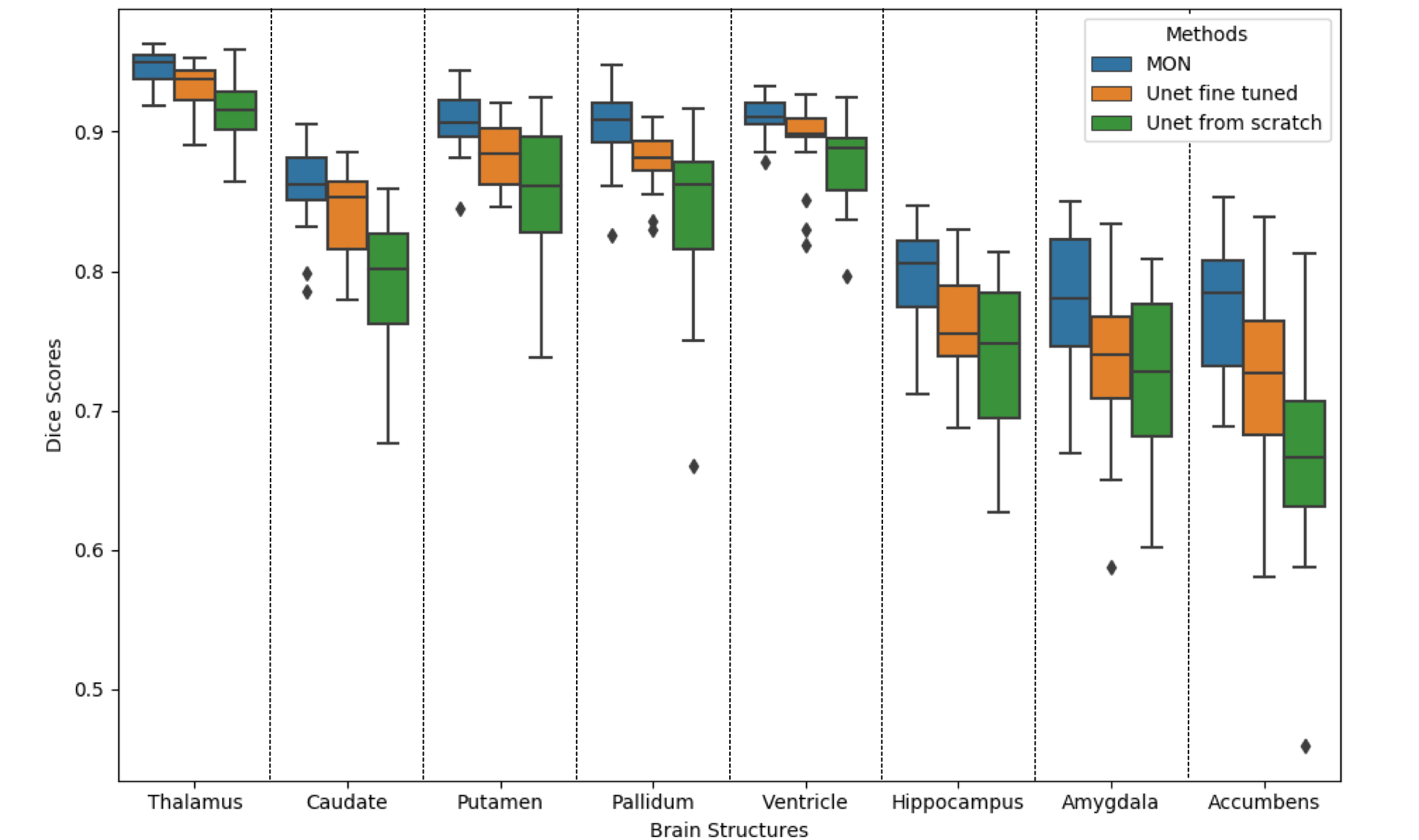}
  \caption{Box-plot of Dice scores of MO-Net, U-Net fine-tuned and U-Net trained from scratch on HAA for 8 brain structures on the left hemisphere.}\label{fig:box}
  \end{center}
\end{figure}

\begin{figure}[h]
\begin{center}
  \includegraphics[scale=0.5]{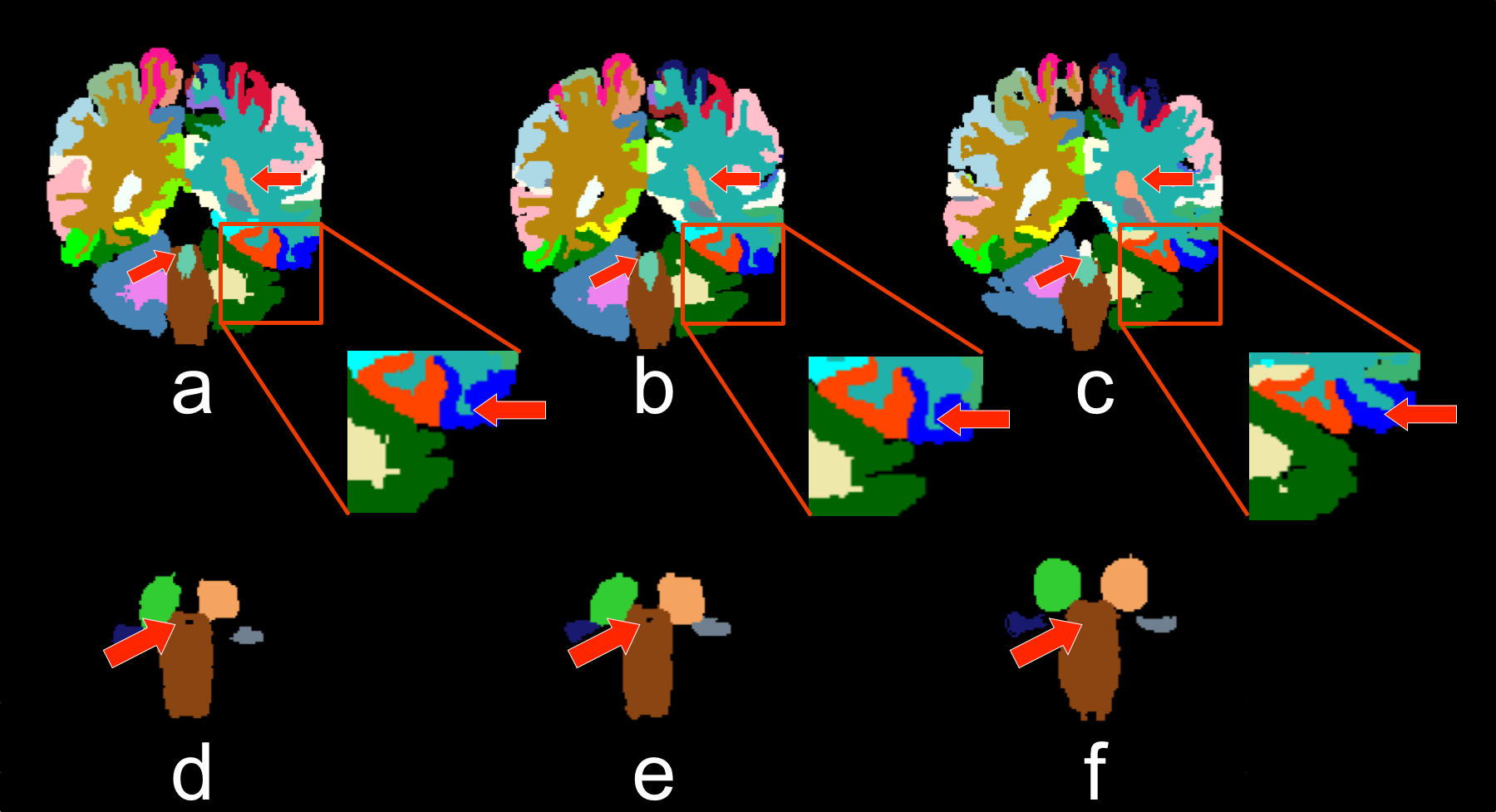}
  \caption{Visual inspection of whole brain segmentation and sub-cortical segmentation on MALC: Ground truth of full (a) and partial (d) brain segmentation from the expert, full (b) and partial (e) brain segmentation from MO-Net, full (c) segmentation from fine-tuned U-Net, and sub-cortical (f) segmentation from FSL. Red arrows indict regions where MO-Net looks consistent with manual annotations and outperforms other methods.}\label{fig:visual}
  \label{sampledir}
  \end{center}
\end{figure}

A box-plot of Dice scores comparing MO-Net with U-Net trained from scratch and fine-tuned on HAA for 8 brain structures is given in Fig.~\ref{fig:box} showing the improvement of adopting our method. A qualitative result of whole brain and sub-cortical segmentation from MO-Net is given in Fig.~\ref{fig:visual}, which shows better segmentation accuracy for certain structures comparing with U-Net and FSL.

The result has demonstrated that a CNN-based model pre-trained with partial segmentation can achieve better accuracy for whole brain segmentation. The performance of MO-Net in terms of Dice scores is comparable to 3D U-Net based approaches in \cite{huo20193d} on MALC with less strict training data, although inferior to \cite{roy2019quicknat} probably due to the deeper network they adopted. We believe the performance of our approach has the potential to be improved with a more advanced CNN design in the future. In general, multi-task learning helps the model to improve the generalization and in our case, to learn features shared by partial segmentation and full segmentation, which can possibly make our encoder more robust. Such claim would need more experiments to prove in the future.

\section{Conclusion}
\label{sec:conclusion}
In this paper, we propose a method that combines transfer learning and multi-task learning to address the small data learning problem. Our method takes advantage of existing automatic tool to create a large set of partial annotations for model pre-training which has been demonstrated to improve segmentation accuracy. The preliminary result on whole brain segmentation shows a good potential of the proposed method.

\section{Acknowledgements}
This research is independent research funded by the NIHR Imperial Biomedical Research Centre (BRC). The views expressed in this publication are those of the author(s) and not necessarily those of the NHS, NIHR or Department of Health. The research is conducted using the UK Biobank Resource under Application Number 18545. We gratefully acknowledge the support of NVIDIA Corporation with the donation of the GPU used for this research.

\bibliographystyle{splncs}
\bibliography{paper44}
% \addbibresource{reference/miccai_ref}
% {\footnotesize\bibliography{IEEEabrv,3DShapeAnalysis}}

\end{document}